\documentclass[11pt,twoside]{article} %TS01-3 %TS01-3 %TS23-2
\usepackage{cspm-asp2006}
\usepackage{epsfig,graphicx,natbib,url}  %RR additions
\usepackage{lscape} %%\usepackage{hyperref}
\begin{document}
\setcounter{page}{239}

\markboth{Bloomfield et al.}{RPW in a Sunspot Chromosphere}
\title{Observations of Running Waves in a Sunspot Chromosphere}
\author{D. Shaun Bloomfield,
        Andreas Lagg
        and 
        Sami K. Solanki}
\affil{MPI f\"{u}r Sonnensystemforschung, Katlenburg-Lindau, Germany}

%%%%%%%%%%%%%%%%%%%%%%%%%%%%%%%%%%%%%%%%%%%%%%%%%%%%%%%%%%%%%%%%%%%%%%%%%%%%
\begin{abstract}
Spectropolarimetric time series data of the primary spot of active
region NOAA 9448 were obtained in the Si\,I\,10827\,\AA\ line
and the He\,I\,10830\,\AA\ multiplet with the Tenerife Infrared
Polarimeter.  Throughout the time series the spectrograph slit was
fixed over a region covering umbra, a light bridge, penumbra, and
quiet sun. We present speeds of running penumbral waves in the
chromosphere, their relation to both photospheric and chromospheric
umbral oscillations, and their dependence on the magnetic field
topology.
\end{abstract}
%%%%%%%%%%%%%%%%%%%%%%%%%%%%%%%%%%%%%%%%%%%%%%%%%%%%%%%%%%%%%%%%%%%%%%%%%%%%

%%%%%%%%%%%%%%%%%%%%%%%%%%%%%%%%%%%%%%%%%%%%%%%%%%%%%%%%%%%%%%%%%%%%%%%%%%%%
\section{Introduction}
%%%%%%%%%%%%%%%%%%%%%%%%%%%%%%%%%%%%%%%%%%%%%%%%%%%%%%%%%%%%%%%%%%%%%%%%%%%%

Running penumbral waves (RPWs), which exist in sunspot chromospheres,
were first observed by \citet{sha-1972ApJ...178L..85Z} and more
recently studied though various imaging observations \citep[e.g., the
series of papers by][]{sha-2000A&A...354..305C,
sha-2000A&A...363..306G, sha-2001A&A...375..617C}.  Although this
phenomenon has been investigated in detail, the origin of these moving
disturbances and their relation to other phenomena occurring within
sunspots remains unclear. In particular, a comparative study by
\citet{sha-2006A&A...456..689T} between the possibility of RPWs being
a trans-sunspot wave in the chromosphere or a visual pattern of
upward-propagating waves was unable to conclude one way or the other.

The work presented here will attempt to finally address the true
origin of these waves using full Stokes time series observations of a
sunspot obtained at high cadence. The benefit of observing the full
Stokes polarization profiles is the retrieval of the complete magnetic
field vector, circumventing the need for any assumptions of the field
topology in our conclusions of possible wave propagation.

%%%%%%%%%%%%%%%%%%%%%%%%%%%%%%%%%%%%%%%%%%%%%%%%%%%%%%%%%%%%%%%%%%%%%%%%%%%%
\section{Observations}
%%%%%%%%%%%%%%%%%%%%%%%%%%%%%%%%%%%%%%%%%%%%%%%%%%%%%%%%%%%%%%%%%%%%%%%%%%%%

The data used here were obtained on 2001 May 9 with the Tenerife
Infrared Polarimeter \citep{sha-1999ASPC..183..264M} attached to the
German VTT. The full Stokes ($I$, $Q$, $U$, $V$) vector was measured
by a slit fixed across active region NOAA 9448. The slit sampled
sunspot umbra, a light bridge, penumbra and neighboring quiet Sun at a
rate of one exposure every 2.1\,s over the period 09:10--10:20~UT. The
recorded data from inside the sunspot umbra were previously presented
by \citet{sha-2006ApJ...640.1153C}, who found vertically propagating,
slow-mode waves in a stratified isothermal atmosphere to reliably fit
the observed temporal Fourier behavior when radiative cooling was
taken into account.

\begin{figure}
  \centering
  \includegraphics[width=12.5cm]{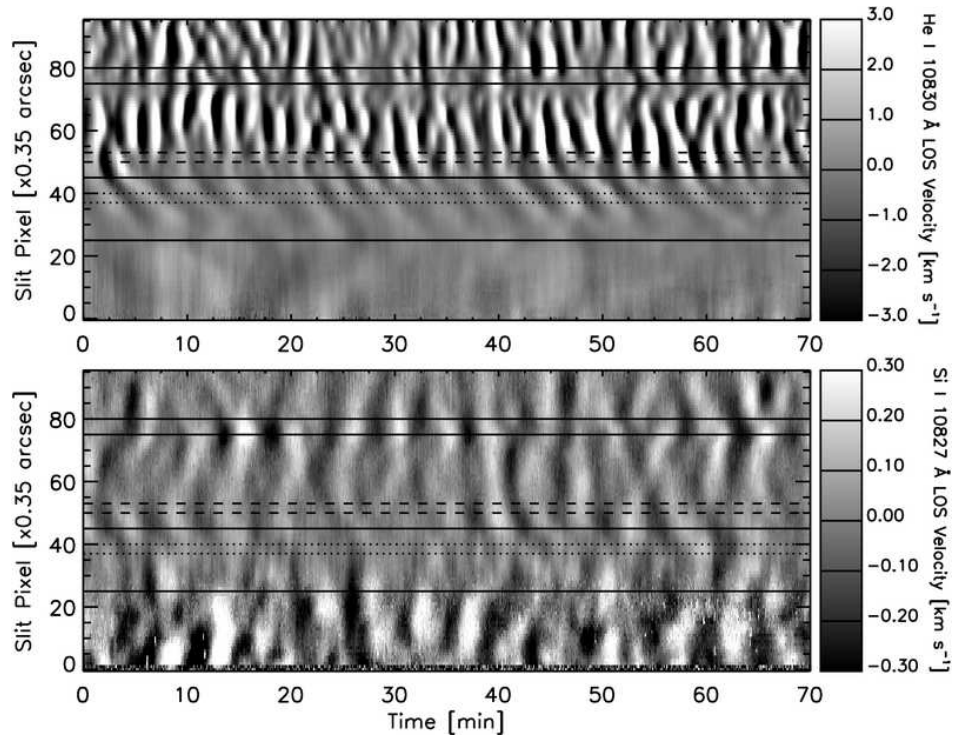}
  \caption[]{\label{bloomfield-fig:si-he-osc}
  Space-time plots of LOS velocity in the photospheric 
  Si\,I\,10827\,\AA\ line (bottom) and upper-chromospheric 
  He\,I\,10830\,\AA\ multiplet (top).
}
\end{figure}

The Stokes profiles of the photospheric Si\,I\,10827\,\AA\ and 
high-chromospheric He\,I\,10830\,\AA\ lines were individually inverted 
using the inversion code of \citet{sha-2004A&A...414.1109L}. Atmospheres 
containing a single magnetic component were used in both cases, while the 
Si\,I inversion included an additional non-magnetic, stray-light 
component. The resulting line-of-sight (LOS) velocities from the inversion 
are displayed in Fig.~\ref{bloomfield-fig:si-he-osc}, where solid 
horizontal lines mark boundaries between different regions of the sunspot: 
quiet Sun ($y = 0-25$); penumbra ($y = 25-45/50$); umbra 
($y = 45/50-95$); a light bridge ($y = 75-80$). It is evident that 
whereas the umbral oscillations in He\,I have a period of 3~min, in 
Si\,I the umbra oscillates at around 5~min. Note, the RPWs also show 
a period close to 5~min.

%===========================================================================
%===========================================================================

%%%%%%%%%%%%%%%%%%%%%%%%%%%%%%%%%%%%%%%%%%%%%%%%%%%%%%%%%%%%%%%%%%%%%%%%%%%%
\section{Horizontal Motions}
%%%%%%%%%%%%%%%%%%%%%%%%%%%%%%%%%%%%%%%%%%%%%%%%%%%%%%%%%%%%%%%%%%%%%%%%%%%%

%===========================================================================
\begin{figure}
  \centering
  \includegraphics[width=\textwidth]{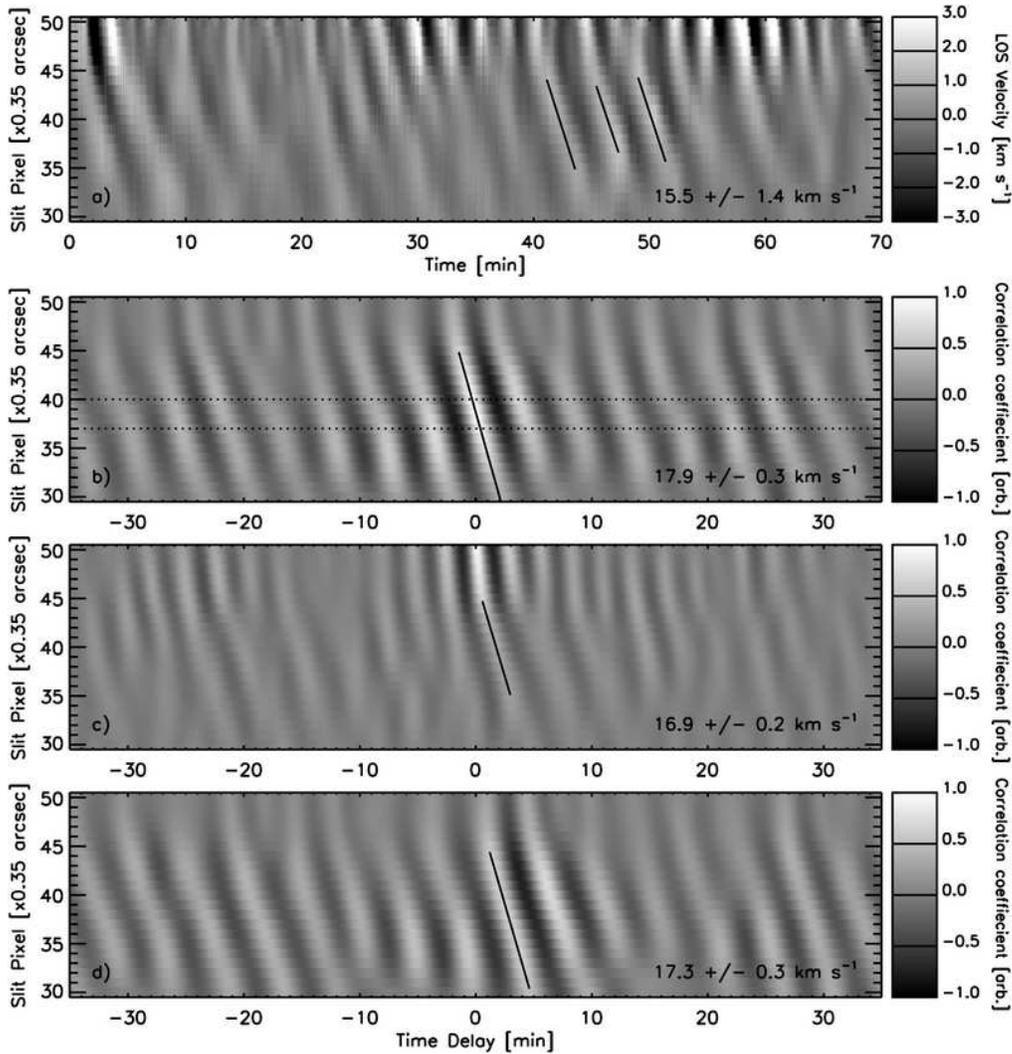}
  \caption[]{\label{bloomfield-fig:raw-he-ccor}
  {{a}}) Space-time plot of penumbral LOS velocity in the He\,I 
  multiplet. Variation of temporal cross correlation between individual 
  spatial He\,I velocity profiles and {{b}}) an average, central 
  penumbral He\,I profile, {{c}}) an average, outer umbral 
  He\,I profile, and {{d}}) an average, outer umbral 
  Si\,I profile.
}
\end{figure}
%===========================================================================

Initially, raw He\,I velocities were studied to determine the apparent
horizontal propagation speed of the RPW disturbances through the
penumbra. Simple spatial tracking of a number of the velocity maxima
in Fig.~\ref{bloomfield-fig:raw-he-ccor}a yielded values
$\approx$$16$\,km\,s$^{-1}$. Time lags between all of the penumbral He\,I
velocity signals and an averaged, mid-penumbral He\,I signal (region
of averaging marked by dotted lines in
Fig.~\ref{bloomfield-fig:si-he-osc}) were found by cross-correlating
the velocity profiles in time and are presented in
Fig.~\ref{bloomfield-fig:raw-he-ccor}b. Spatial tracking of
the peak correlation yielded an apparent speed of
$\approx$$18$\,km\,s$^{-1}$. The same spatial correlation approach was
also performed between all of the penumbral He\,I velocity signals and
both an averaged, outer-umbral He\,I signal
(Fig.~\ref{bloomfield-fig:raw-he-ccor}c; region of averaging
marked by dashed lines in Fig.~\ref{bloomfield-fig:si-he-osc}) and an
averaged, outer-umbral Si\,I signal
(Fig.~\ref{bloomfield-fig:raw-he-ccor}d), each yielding
apparent speeds of $\approx$$17$\,km\,s$^{-1}$. These values are in good
agreement with the previous literature \citep[see, e.g., the recent
review of][]{sha-2006RSPTA.364..313B}. Note that values of correlation
coefficient in the umbral He\,I case diminish rapidly through the
penumbra since attempting to correlate a signal with a dominant 3-min
period to one with a 5-min period. The correlation with the umbral
Si\,I displays the opposite behavior, being strong throughout the
penumbra but petering out at the umbra/penumbra boundary for the same
reason.

%%%%%%%%%%%%%%%%%%%%%%%%%%%%%%%%%%%%%%%%%%%%%%%%%%%%%%%%%%%%%%%%%%%%%%%%%%%%
\section{Upward Propagation}
%%%%%%%%%%%%%%%%%%%%%%%%%%%%%%%%%%%%%%%%%%%%%%%%%%%%%%%%%%%%%%%%%%%%%%%%%%%%

\begin{figure}
  \centering
  \includegraphics[width=\textwidth]{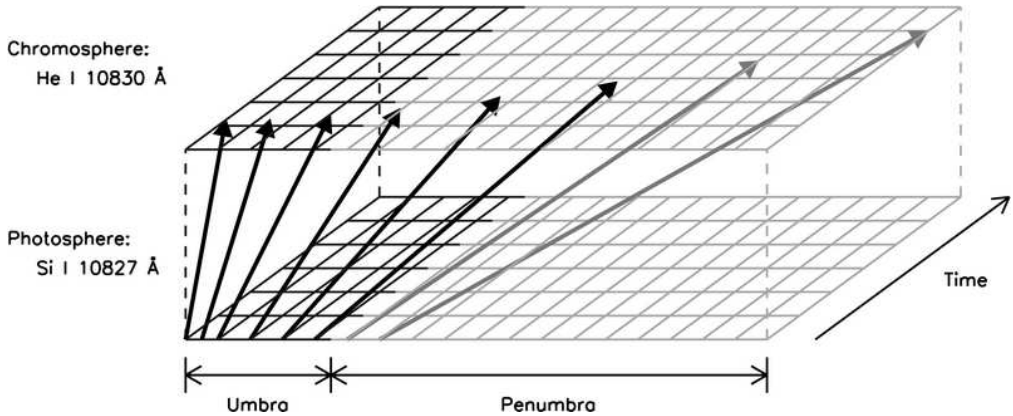}
  \caption[]{\label{bloomfield-fig:car-sch-inc}
  Cartoon schematic of the proposed field geometry linking pixels in the umbra 
  (dark grid) to those in the penumbra (light grid) at photospheric and 
  chromospheric sampling heights through increasingly inclined field lines.
}
\end{figure}

The scenario proposed here for the generation of these running
disturbances in the penumbral chromosphere is shown in
Fig.~\ref{bloomfield-fig:car-sch-inc} as a cartoon schematic. Although
assuming vertical fields in the umbra of this sunspot was sufficient
for the work of \citet{sha-2006ApJ...640.1153C}, pixels toward the
outer edge of the umbra show somewhat inclined fields
\citep[$35-45^{\circ}$; which may account for the spread of phase
difference points with frequency in][]{sha-2006ApJ...640.1153C}. These
inclinations mean that some field lines originating from the
photospheric umbra can pass through the chromospheric penumbra. From
the values of inclination retrieved here, most chromospheric penumbral
pixels can be traced back to the outer umbra/inner penumbra, all of
which show photospheric velocity signatures of similar phase. If waves
which travel along the field are indeed excited at the photospheric
level within the umbral region \citep[as shown
by][]{sha-2006ApJ...640.1153C}, those waves propagating along more
inclined paths will have larger distances to traverse before showing
their presence at the He\,I sampling height in the upper
chromosphere. The increased path lengths that are experienced with
increasing distance through the chromospheric penumbra mean that
initially similar velocity signals in the Si\,I line have increasing
temporal delay in the He\,I signal, resulting in an apparent
horizontal motion outward from the umbra as proposed by
\citet{sha-2003A&A...403..277R}.

%===========================================================================
%===========================================================================

\begin{figure}
  \centering
  \includegraphics[width=12.5cm]{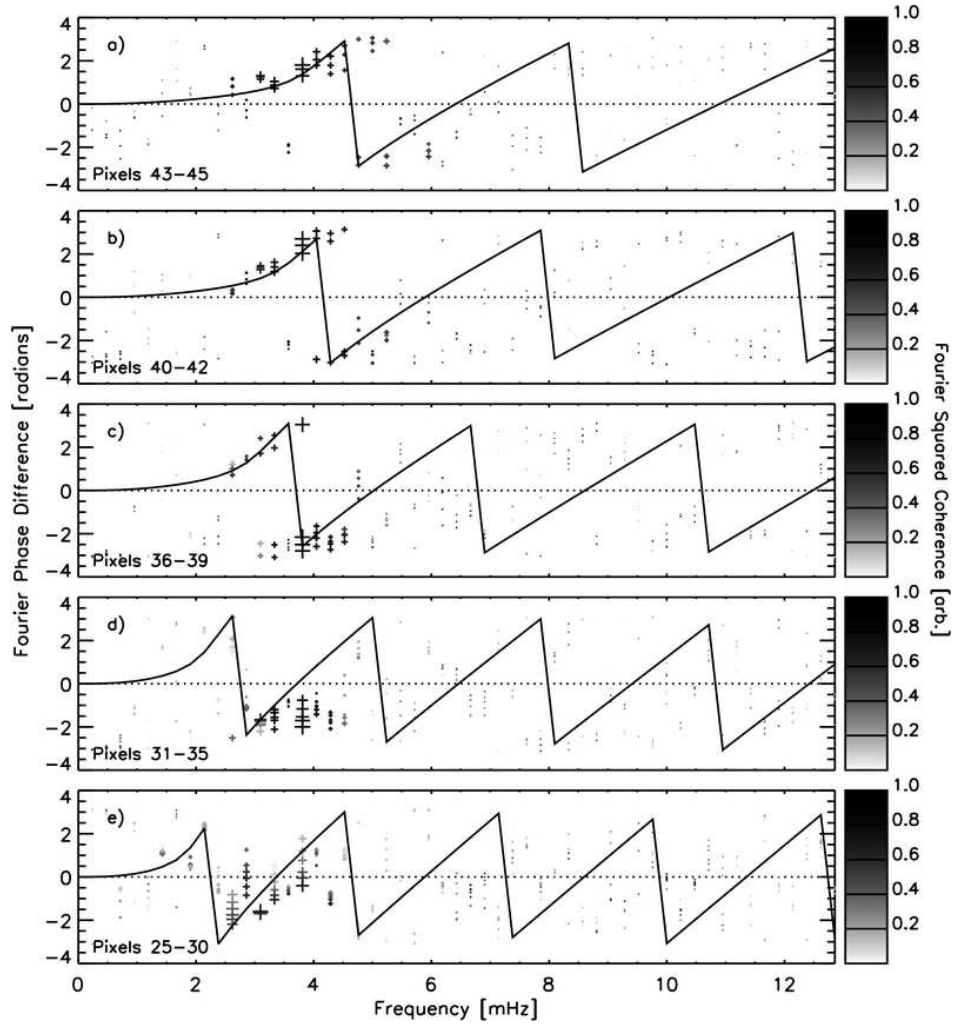}
  \caption[]{\label{bloomfield-fig:fft-ang-dep}
  Phase differences between spatially offset pairings of photospheric and 
  chromospheric LOS velocities. Panels show phase difference relations 
  moving from those chromospheric pixels closest to the umbra/penumbra 
  boundary (top) toward those at the penumbra/quiet-sun boundary 
  (bottom).
}
\end{figure}

Photospheric and chromospheric spatial pixels were paired using the
values of local solar inclination recorded in the Si\,I and He\,I
lines to determine the expected spatial offset between the two
sampling heights in the atmosphere. The subsequent LOS velocity
pairings between the photosphere and chromosphere were subjected to
Fourier phase difference analysis following
\citet{sha-2001A&A...379.1052K}. Phase difference spectra from several
neighboring spatial pixels were overplotted in each of the panels of
Fig.~\ref{bloomfield-fig:fft-ang-dep} to heighten the clarity of any
relation which may exist. These phase difference diagrams are
comparable to Fig.~6 of \citet{sha-2006ApJ...640.1153C}, although they
differ slightly as the shade and size of data points here denote the
Fourier coherence and cross-spectral power, respectively.

%===========================================================================
%===========================================================================

It is clear when moving from the chromospheric umbra/penumbra boundary
(top panel) toward the penumbra/quiet sun boundary (bottom panel) that
the phase difference relations show deviation from the classical
acoustic cutoff ($\approx 5.2$~mHz) which can be explained by the
reduced gravity experienced by waves propagating non-vertically. Solid
curves plotted in Fig.~\ref{bloomfield-fig:fft-ang-dep} show the
expected phase difference relations achieved using the equations
provided by \citet{sha-2006ApJ...640.1153C} for field-aligned waves in
an isothermal atmosphere with radiative cooling, when the reduced
gravity and increased path length from inclinated fields are taken
into account, and the values of temperature, height separation, and
cooling time scale which they determined for the umbra of this sunspot
($T=4000$\,K, $\Delta z=1000$\,km, and $\tau_{\rm{R}}=55$\,s,
respectively). Although only arbitrarily chosen and not reached by any
form of fitting, good comparison is found with the observations for
field inclinations of $40^{\circ}$, $45^{\circ}$, 53$^{\circ}$,
63$^{\circ}$, and 65$^{\circ}$ when moving from top to bottom panels
of Fig.~\ref{bloomfield-fig:fft-ang-dep}: these values are also in
good agreement with the local solar inclinations determined by the
Stokes inversion.

%%%%%%%%%%%%%%%%%%%%%%%%%%%%%%%%%%%%%%%%%%%%%%%%%%%%%%%%%%%%%%%%%%%%%%%%%%%%
\section{Future Work}
%%%%%%%%%%%%%%%%%%%%%%%%%%%%%%%%%%%%%%%%%%%%%%%%%%%%%%%%%%%%%%%%%%%%%%%%%%%%

Given that the work presented here is incomplete, the initial results
are very promising. Fourier analysis confirms that RPWs are indeed the
visual pattern of low-$\beta$ slow magneto-acoustic waves that are
generated at very
similar phase at the photosphere but propagate along field lines of
increasing inclination, hence showing increased time delays at
(roughly) the same chromospheric altitude.

Improvements that are still required include least-squares fitting of the 
phase difference relation to the data to retrieve the spatial variation of 
the physical parameters $T$, $\Delta z$, and $\tau_{\rm{R}}$. In addition, 
refinement of the Si\,I inversion to include fine structure in the 
photospheric penumbra (e.g., a second magnetic component to account for 
known highly inclined filament or flux tube fields) will provide better 
linkage between velocity signals in the inner-penumbral photosphere and 
those in the outer-penumbral chromosphere. Observations of a small pore 
region will also be investigated for such running waves since the findings 
of this work indicate that the existence of a penumbra is not neccessary for 
their production: only sufficiently inclined fields in the umbra (or pore) 
are required to guide the waves outside this region.\\

%%%%%%%%%%%%%%%%%%%%%%%%%%%%%%%%%%%%%%%%%%%%%%%%%%%%%%%%%%%%%%%%%%%%%%%%%%%%
\acknowledgements 
The authors wish to extend their thanks to Rebecca Centeno, Manuel Collados, 
and Javier Trujillo Bueno for providing this excellent data set for our 
analysis. We are also grateful to the conference organisers for the 
opportunity to upgrade this presentation from a poster to a talk.

%% References via BibTeX
%%%%%%%%%%%%%%%%%%%%%%%%%%%%%%%%%%%%%%%%%%%%%%%%%%%%%%%%%%%%%%%%%%%%%%%%%%%%
%RR Use the following two commands to generate the bibliography
%RR automatically with BibTeX, and then insert the resulting .bbl file,
%RR or collect \bibitem info for each paper from ADS (``preferred format'')

\bibliographystyle{cspm-bib}		%RR copy of my old aabib.bst
%\bibliography{bloomfield-adsfiles}	%## your own ads-collected .bib file(s)

\end{document}